# Machine Learning Automatically Detects COVID-19 using Chest CTs in a Large Multicenter Cohort

**Authors:**

Eduardo Jose Mortani Barbosa Jr., MD (6)*, Bogdan Georgescu, PhD (8)*, Shikha Chaganti, PhD (8), Gorka Bastarrika Aleman, MD (1), , Jordi Broncano Cabrero, MD (2), Guillaume Chabin (9), Brian Teixeira (8), Thomas Flohr, apl. Prof., PhD (7), Philippe Grenier, Prof., MD (3), Sasa Grbic, PhD (8), Nakul Gupta, MD (4), François Mellot, MD (3), Savvas Nicolaou, MD (11), Thomas Re, MD (8), Pina Sanelli, MD (5), Alexander W. Sauter, MD (10), Youngjin Yoo, PhD (8), Valentin Ziebandt (7), Dorin Comaniciu, PhD (8)

(1) Clínica Universidad de Navarra, Navarra, Spain
(2) Health Time, Jaén, Spain
(3) Hôpital Foch, Suresnes, France
(4) Houston Methodist, Houston, USA
(5) Donald and Barbara Zucker School of Medicine, Feinstein Institutes for Medical Research, Northwell Health, Manhasset, NY, USA
(6) Penn Medicine, Philadelphia, PA, USA
(7) Siemens Healthineers, Forchheim, Germany
(8) Siemens Healthineers, Princeton, NJ, USA
(9) Siemens Healthineers, Paris, France
(10) University Hospital Basel, Clinic of Radiology & Nuclear medicine, Basel, Switzerland
(11) Vancouver General Hospital, Vancouver, Canada

* These authors contributed equally to this manuscript

**Abstract**

Objectives: To investigate machine-learning classifiers and interpretable models using chest CT for detection of COVID-19 and differentiation from other pneumonias, ILD and normal CTs.

Methods: Our retrospective multi-institutional study obtained 2096 chest CTs from 16 institutions (including 1077 COVID-19 patients). Training/testing cohorts included 927/100 COVID-19, 388/33 ILD, 189/33 other pneumonias, and 559/34 normal (no pathologies) CTs. A metric-based approach for classification of COVID-19 used interpretable features, relying on logistic regression and random forests. A deep learning-based classifier differentiated COVID-19 via 3D features extracted directly from CT attenuation and probability distribution of airspace opacities.

Results: Most discriminative features of COVID-19 are percentage of airspace opacity and peripheral and basal predominant opacities, concordant with the typical characterization of COVID-19 in the literature. Unsupervised hierarchical clustering compares feature distribution across COVID-19 and control cohorts. The metrics-based classifier achieved AUC=0.83, sensitivity=0.74, and specificity=0.79 of versus respectively 0.93, 0.90, and 0.83 for the DL-based classifier. Most of ambiguity comes from non-COVID-19 pneumonia with manifestations that overlap with COVID-19, as well as mild COVID-19 cases. Non-COVID-19 classification performance is 91% for ILD, 64% for other pneumonias and 94% for no pathologies, which demonstrates the robustness of our method against different compositions of control groups.

Conclusions: Our new method accurately discriminates COVID-19 from other types of pneumonia, ILD, and no pathologies CTs, using quantitative imaging features derived from chest CT, while balancing interpretability of results and classification performance, and therefore may be useful to facilitate diagnosis of COVID-19.

**Key Points:**

- Unsupervised clustering reveals the key tomographic features including percent airspace opacity and peripheral and basilar opacities most typical of COVID-19 relative to control groups.

- COVID-19 positive were compared with COVID-19 negative chest CTs (including a balanced distribution of non-COVID-19 pneumonia, interstitial lung disease, and no pathologies). Classification accuracies for COVID-19, pneumonia, ILD, and no pathologies CT scans are respectively 90%, 64%, 91%, and 94%.

- Our deep learning (DL)-based classification method demonstrates AUC of 0.93 (sensitivity 90%, specificity 83%). Machine learning methods applied to quantitative chest CT metrics can therefore improve diagnostic accuracy in suspected COVID-19, particularly in resource constrained environments.

**Abbreviations:**

- **COVID-19:** Coronavirus Disease 2019
- **SARS-CoV-2:** Severe Acute Respiratory Syndrome Coronavirus 2
- **RT-PCR:** Reverse Transcript Polymerase Chain Reaction
- **GGO:** Ground Glass Opacity
- **DL:** Deep Learning
- **ILD:** Interstitial Lung Disease
- **GBT:** Gradient Boosted Trees
- **LR:** Logistic Regression
- **RF:** Random Forest
- **PO:** Percent of Opacity
- **PHO:** Percent of High Opacity
- **CO-RADS:** COVID-19 Reporting and Data System
- **PACS:** Picture Archiving and Communication System
- **ROC:** Receiver Operating Characteristic

- **AUC:** Area Under the Curve
- **DICOM:** Digital Imaging and Communications in Medicine

**Introduction**

Coronavirus disease 2019 or COVID-19 has caused a global pandemic associated with an immense human toll and healthcare burden across the world (1). COVID-19 can manifest as pneumonia, which may lead to acute hypoxemic respiratory failure, the main reason for hospitalization and mortality. A Fleischner Society statement supports the use of lung imaging for differential diagnosis and management of patients with moderate to severe clinical symptoms, especially in resource-constrained environments (2). The most typical pulmonary CT imaging features of COVID-19 are multifocal (often bilateral and peripheral predominant) airspace opacities, comprised by ground-glass opacities and/or consolidation, which may be associated with interlobular and intralobular septal thickening (3). A study comparing COVID-19 and other types of viral pneumonia demonstrated that distinguishing features more typical of COVID-19 are predominance of ground-glass opacities, peripheral distribution, and perivascular thickening (4). A consensus statement on COVID-19 reporting by Radiological Society of North America indicates the typical appearance of COVID-19 as peripheral and bilateral distribution of ground-glass opacities with or without consolidation, and possibly with the 'reverse halo' sign (5). Confirmatory diagnosis of COVID-19 requires identification of the virus on nasopharyngeal swabs via RT-PCR (reverse transcription - polymerase chain reaction), a highly specific test (>99%) but with lower sensitivity (50-80%) (6,7). Given the imperfect sensitivity of RT-PCR and potential resource constraints, chest CT imaging has an evolving role in diagnosis of COVID-19, and possibly prognostic value.

Recently, several groups have shown that COVID-19 can be identified on CT with variable accuracy. For example, chest CTs in patients who were positive for COVID-19 (RT-PCR) could be distinguished from chest CTs in patients that tested negative with an AUC of 0.92 using machine learning (8). While this classification is potentially valuable, it is limited by lack of details on the types and distribution of findings on negative (control) cases. It is important to distinguish COVID-19 related pulmonary disease not just from subjects with no pathologies CTs, but also from other types of lung diseases unrelated to COVID-19, including other infections, malignancy, ILD and COPD. This is especially important as COVID-19 can manifest similarly clinically to other respiratory infections such as influenza, which can lead to confusion in triage and diagnosis. Bai et al. showed that an artificial intelligence system can assist radiologists to distinguish COVID-19 from other types of pneumonia, with diagnostic sensitivity to 88% and specificity to 90% (9). The two cohorts (COVID-19 and other pneumonia) compared in this study are from two different countries, limiting the generalizability of their model. Other studies showing promising results in classification do not provide a detailed description of imaging cohorts acquisition protocols or data sources (10,11). This information is important since different institutions will have diverse CT acquisition protocols and clinical indications for CT usage, which can affect the performance of machine learning algorithms.

Our goals were: compute CT derived quantitative imaging metrics corresponding to the typical presentation of COVID-19 and evaluate the discriminative power of these metrics for the diagnosis of COVID-19, when compared to different compositions of control groups; perform unsupervised clustering of interpretable features to visualize how COVID-19 patients differ from controls; compare the performance of metrics-based classifiers to a deep learning-based model. Our large training and test datasets contained chest CTs in patienst confirmed with COVID-19 and negative controls from multiple institutions in North America and Europe, making this one of the first large studies to demonstrate the value of machine learning for differentiation of COVID-19 and non-COVID-19 utilizing data from multiple centers, increasing generalizability and applicability.

**Methods**

*Patient Selection and Imaging Data:*

This retrospective study utilized data acquired from 16 different centers in North America and Europe, after anonymization, ethical review and approval at each institution. Our dataset consists of chest CTs of 1226 patients positive for COVID-19, and 1287 chest CTs of patients without COVID-19, including with other types of pneumonia (n=240), interstitial lung disease (ILD) (n=437), and without any pathologies on chest CT (n=610). The flowchart for patient selection criteria is shown in Figure 1. All CTs in the COVID-19 cohort from North America have been confirmed by RT-PCR. The COVID-19 cohort from Europe has been either confirmed by RT-PCR or diagnosed based on clinical symptoms, epidemiological exposure and radiological assessment. The pneumonia cohort consists of cases of patients with non-COVID-19 viral or bacterial pneumonias, organizing pneumonia and aspiration pneumonia. The ILD cohort consists of patients with various types of ILD such as usual interstitial pneumonia, nonspecific interstitial pneumonia, and other unclassifiable interstitial diseases with our without fibrotic features, which exhibit ground-glass opacities, reticulation, honeycombing, traction bronchiectasis and consolidation to different degrees. 64 COVID-19 cases were excluded due to no opacities on Chest CT, 84 COVID-19 cases were excluded due to had minimal opacities (PO<1%), one COVID-19 case was excluded due to incomplete inclusion of the lungs in the field-of-view, two pneumonia controls were excluded due to incorrect DICOM parameters and imaging artifacts.

The dataset was divided into training (2063), validation (99) and test (200) sets (Table 1). Model training and selection was performed based on training and validation sets. The final performance of selected models is reported on the test dataset (Table S1 provides detailed breakdown of demographic and scanning information for each cohort).

*Metrics of Airspace Disease Severity*

We computed thirty-two metrics of severity based on abnormalities known to be associated with COVID-19, as well as lung and lobar segmentation, using a previously developed Deep Image-to-Image Network trained on a large cohort of healthy and abnormal cases for lung segmentation (12). Next, we used a DenseUnet to identify abnormalities such as GGO and consolidations on COVID-19 as well as control groups (12). Based on these segmentations, we computed severity metrics to summarize the spatial distribution and extent of airspace disease in both lungs. Complete and detailed description of the thirty-two metrics is provided in Table S2. Our algorithm is fully automated and requires no manual input (Figure 2).

*Metric-based Analysis*

*Unsupervised Feature Selection and Clustering*

Recursive feature elimination was used to select the metrics of severity most discriminative between COVID-19 and non-COVID-19 classes. The $k$ best features were selected based on an internal validation split. Based on the selected metrics, an unsupervised hierarchical cluster analysis was performed to identify clusters of images that have similar features. The pairwise Euclidean distance between two metrics was used to compute a distance matrix, with average linkage method used for hierarchical clustering (13), visualized as a heatmap (Figure 3).

*Supervised COVID-19 Classification*

Two metrics-based classifiers were trained based on the thirty-two computed metrics. First, we trained a Random Forest classifier, M1, using $k$ selected features based on recursive feature elimination. Subsequently, we trained a second classifier using logistic regression (LR), after feature transformation based on gradient boosted trees (GBT) (14). For training GBT, we used 2000 estimators with max depth=3 and 3 features for each split. The boosting fraction 0.8 was used for fitting the individual trees. The Logistic Regression classifier, M2, was trained with L2 regularization (C=0.2). The class weights were adjusted to class frequencies for the class imbalance between COVID-19 and non-COVID-19 classes.

*Supervised Deep-learning based COVID-19 classification*

A deep-learning-based 3D neural network model, M3, was trained to separate the positive class (COVID-19) vs negative class (non-COVID-19). As input, we considered a two-channel 3D tensor, with the first channel containing directly the CT Hounsfield units within the lung segmentation masks and the second channel containing the probability map of a previously proposed opacity classifier (12). The 3D network uses anisotropic 3D kernels to balance resolution and speed with deep dense blocks that gradually aggregate features down to a binary output. The network was trained end-to-end as a classification system using binary cross entropy and uses probabilistic sampling of the training data to adjust for the imbalance in the training dataset labels. A separate validation dataset was used for final model selection before the performance was measured on the testing set. The input 3D tensor size is fixed (2x128x384x384) corresponding to the lung segmentation from the CT data rescaled to a 3x1x1mm resolution. The first two blocks are anisotropic and consist of convolution (kernels 1x3x3) – batch normalization – LeakyReLU and Max-pooling (kernels 1x2x2, stride 1x2x2). The subsequent five blocks are isotropic with convolution (kernels 3x3x3) – batch normalization – LeakyReLU and Max-pooling (kernels 2x2x2, stride 2x2x2) followed by a final linear classifier with the input 144-dimensional. Figure 2 depicts our 3D DL classifier.

*Comparison with models from the literature*

We compared the models in this work to those published by Li et al(10) and the Clara model proposed by Harmon et al (15). Li et al (10) investigated a deep learning method to distinguish COVID-19 from community-acquired pneumonia and healthy subjects using chest CT. Their proposed DL method is based on extracting 2D features on each CT slice followed by feature pooling across slices and a final linear classifier. There are two main differences between the DL method proposed in this article and the one proposed by Li et al (10). First, our method is based on 3D deep learning, which better leverages the contiguity of imaging textures along the z-axis, and second, it uses as input the spatial distribution of opacities within the lung parenchyma, which focuses the classifier on the regions of abnormality. For the method of Li et al (10) we have re-trained the model on our dataset while for Harmon et al (15) we have run their released model on our testing set.

**Results**

Six features were selected by recursive feature elimination between features and classes in the training dataset of 927 COVID-19 cases and 1136 controls (pneumonia, ILD and no pathologies). The features are:

1) Percent of Ground Glass Opacities
2) Percent of Opacity (PO) (consolidation and ground-glass opacities=airspace disease)
3) Percent of Opacities in the Periphery (see appendix)
4) Percent of Opacities in the Rind (see appendix)
5) Percent of Opacities in the Right Lower Lobe
6) Percent of Opacities in the Left Lower Lobe

These features correspond to reported typical COVID-19 features: multifocal ground-glass opacities and/or consolidation with basilar and peripheral distribution (3) (5) (16). Figure 3 demonstrates the hierarchical clustering of these metrics, along with the ground-truth diagnosis cohort membership (COVID-19, other pneumonia, ILD and no pathologies CT) shown on the band on the left of heat map. The metric values are standardized and rescaled to a value between 0 and 1. In Figure 3(a), the clustering is performed on the entire training set of 1800 subjects. The probability of belonging to the COVID-19 class increases towards the bottom of the heat map, which corresponds to higher values of the metrics, i.e., more opacities (both GGO and consolidation), and more peripheral and basilar distribution. The middle of the heatmap shows the ambiguous region, where there is an overlap of features from different disease cohorts. Figure 3(b) shows the same clustering in the test dataset for each of the disease cohorts. While there is a cluster of COVID-19 subjects that have characteristic features, there are also many which do not show all characteristics. Moreover, some cases of pneumonia and ILD overlap with the typical features of COVID-19.

The six selected features were used to train a random forest classifier (M1). The performance of this classifier on a test dataset has an AUC of 0.75 (95% CI: [0.69, 0.81]) as depicted in Figure 4, which shows bootstrapped ROC and AUC values, along with their 95% confidence intervals, which were computed on 1000 samples with replacement. The sensitivity and specificity of this model are 0.86 and 0.60, respectively. The performance is improved by training a second classifier on all thirty-two metrics using a logistic regression model (M2). The metrics are first transformed to a higher-dimensional space using feature embedding with gradient boosted trees. On the test set this model produces an AUC of 0.83 (95% CI: [0.78, 0.89]) with a sensitivity of 0.74 and a specificity of 0.79. While the performance improves, some of the interpretability is lost since the features are transformed to a higher dimension.

Our deep learning-based classifier (M3) has the best performance with an AUC of 0.93 (95% CI: [0.90, 0.96]), improving the sensitivity and specificity of the system to 0.90 and 0.83, respectively. The improvement is mostly due to a reduction of false positives from the ILD and increase of true positives in the COVID-19 class. The optimal operating point for all models was chosen as the point with the shortest distance from the top left corner on the ROC computed on the whole test dataset, without bootstrapping (17). The corresponding confusion matrices for the three models are shown in Table 2. Figure 5 illustrates examples of correctly labeled samples by the metrics-based classifier and the DL-based classifier, on typical CT images from COVID-19 patients. Figure S2 shows negative examples from ILD and non-COVID-19 pneumonia patients. Overlaid in red are the areas identified by the opacity classifier. Figure 6 illustrates examples of cases incorrectly labeled by both classifiers and Figure S3 shows cases that are incorrectly labeled by the metric-based classifier but correctly labeled by the DL classifier that uses additional texture features extracted directly from the images.

We compared the models in this work to those published by Li et al(10) and the Clara model proposed by Harmon et al (15). We trained and tested on our dataset using the published code by Li et al (10) and achieved an AUC of 0.90 (95% CI: [0.86, 0.94]) as shown in Figure S2. For Harmon et al (15) we have run their released model on our testing set and achieved an AUC of 0.74 (95% CI: [0.67, 0.81]). The optimal operating point, which was selected as the point closest to the top left corner of the ROC computed on the whole test dataset, without bootstrapping, produced a sensitivity of 0.86 and specificity of 0.80 for Li et al and a sensitivity of 0.64 and specificity of 0.78 for Harmon et al (Figure 4b). The confusion matrix is shown in Table 3.

**Discussion**

We evaluated the ability of machine learning algorithms to distinguish between chest CTs in patients with COVID-19 and a diverse control cohort comprising of chest CTs demonstrating other pneumonias, ILD and no pathology. We performed an analysis based on clinically interpretable severity metrics computed from automated segmentation of abnormal regions in a chest CT scan, as well as using a deep learning system. Unsupervised clustering on selected severity metrics shows that while there are dominant characteristics that can be observed in COVID-19 such as ground-glass opacities as well as peripheral and basal distribution, these are not observed in all cases of COVID-19. On the other hand, some ILD and other pneumonia patients can exhibit similar characteristics. We found that the performance of the system can be improved by mapping these metrics into a higher dimensional space prior to training a classifier, as shown by model M2 in Figure 3. The best classification accuracy is achieved by the deep learning system, which is essentially a high-dimensional, non-linear model.

The deep learning method achieves reduced false positive and false negative rates relative to the metrics-based classifier suggesting that there might be other latent radiological manifestations of COVID-19 that distinguish it from ILDs or other types of pneumonia. It is worthy investigating how to incorporate the common imaging features into our 3D DL classifier as prior information. The proposed AI-based method has been trained and tested on a database of 2362 CT datasets with 1077 COVID-19 patients and 1285 datasets coming from other categories. We also show how our method compares to the one published by Li et al (10) and found that our method achieves a higher AUC as well as sensitivity. Further details are provided in the supplementary section.

One limitation of this study is that our training set is biased toward COVID-19 and normal controls, potentially affecting discrimination of other lung pathologies. Another limitation is that the validation set size is relatively small, which might not capture the entire data distribution of clinical use cases for optimal model selection. Among the strengths of this study are the diversity of training and testing CTs used, acquired using a variety of CT scanners, as well as numerous institutions and regions, ensuring that robust and generalizable results. We also included not only normal controls, but also various types of lung pathology in the COVID-19 negative group.

Our system provides clinical value in several aspects: it can be used for rapid triage of positive cases, particularly in resource constrained environments where radiologic expertise or RT-PCR may not be immediately available; It could help radiologists to prioritize interpreting CTs in patients with COVID-19 by screening out lower probability cases. The output of our deep learning classifier is easily reproducible and replicable, mitigating inter-reader variability. While RT-PCR will remain the reference standard for confirmatory diagnosis of COVID-19, machine learning methods applied to quantitative CT can perform with high diagnostic accuracy, increasing the value of imaging in diagnosis and management of this disease.

Furthermore, these algorithms could be integrated in a surveillance effort for COVID-19, even in unsuspected patients, in high incidence regions, with automatic assessment for evidence of COVID-19 lung disease, allowing more rapid institution of isolation protocols. Finally, it could potentially be applied retrospectively to large numbers of chest CT exams from institutional PACS systems worldwide to uncover the dissemination of SARS-CoV-2 in communities prior to the implementation of widespread testing efforts.

In the future, we plan to deploy and validate the algorithm in a clinical setting and evaluate the clinical utility and diagnostic accuracy on prospective data, as well as to investigate the value to predict clinical severity and prognosis of COVID-19, as well as ancillary findings of COVID-19 such as acute pulmonary embolism, which is associated with severe COVID-19 (18,19). In addition, clinical decision models could be improved by training a classifier that incorporates other clinical data such as pulse oximetry and laboratory metrics, in addition to imaging features.

**Acknowledgements**

We gratefully acknowledge the contributions of multiple frontline hospitals to this collaboration. The authors also thank the COPDGene for providing the data. The COPDGene study (NCT00608764) was funded by NHLBI U01 HL089897 and U01 HL089856 and also supported by the COPD Foundation through contributions made to an Industry Advisory Committee comprised of AstraZeneca, BoehringerIngelheim, GlaxoSmithKline, Novartis, and Sunovion. We thank many colleagues who made this work possible in a short amount of time. Special recognition to Sebastien Piat, who was instrumental for implementing and managing the data infrastructure.

**TABLES and FIGURES**

Table 1 - Data-split table by classes and categories into training, validation and test datasets.

| 2 classes | 4 categories | Train | Validation | Test |
|---|---|---|---|---|
| Positive | COVID-19 | 927 | 50 | 100 |
| Negative | Pneumonia | 189 | 16 | 33 |
|  | ILD | 388 | 16 | 33 |
|  | No pathology | 559 | 17 | 34 |

Table 2 - Metrics based classifier confusion matrices. The models were evaluated with 100 COVID-19 positive, 33 ILD, 33 pneumonia and 34 no pathologies CT scans. The operating point was chosen as the closest point to the top left corner on the ROC computed over the test dataset (without bootstrapping). NOTE: The table shows the Prediction vs ground truth for each of the negative class CATEGORIES (ILD, other pneumonia, no pathology)

|  |  | Ground Truth | | | |
|---|---|---|---|---|---|
|  |  | Positive | Negative | | |
|  |  | COVID-19 | ILD | Pneumonia | Healthy |
| **Predicted (M1)** | Positive | **86** | 21 | 19 | 0 |
|  | Negative | 14 | **12** | **14** | **34** |
| **Predicted (M2)** | Positive | **74** | 11 | 10 | 0 |
|  | Negative | 26 | **22** | **23** | **34** |
| **Predicted (M3)** | Positive | **90** | 3 | 12 | 2 |
|  | Negative | 10 | **30** | **21** | **32** |

Table 3 - Confusion matrix for the model from Li et al and Harmon et al. The operating point was chosen as the closest point to the top left corner on the ROC computed over the test dataset (without bootstrapping).

|  |  | GT |  |  |  |
| --- | --- | --- | --- | --- | --- |
|  |  | Positive | Negative | | |
|  |  | COVID-19 | ILD | Pneumonia | Healthy |
| Predicted (Li et al) | Positive | **86** | 6 | 14 | 0 |
|  | Negative | 14 | **27** | 19 | **34** |
| Predicted (Harmon et al) | Positive | 64 | **14** | 7 | 1 |
|  | Negative | 36 | **19** | 26 | **33** |

Figure 1. Selection criteria for the COVID-19 and control cohorts in the study

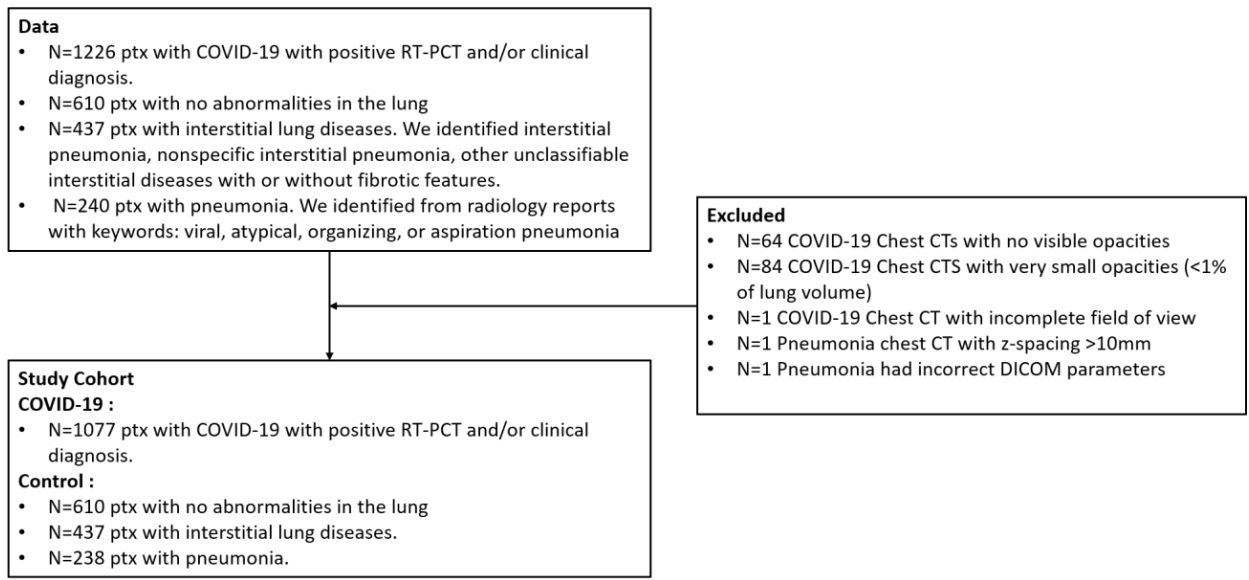

Figure 2. Overview of the deep learning based COVID-19 classifier. Preprocessing consists of lung segmentation and opacities probability distribution computation (12) followed by a 3D deep neural network trained to distinguish between the COVID-19 class and nonCOVID-19 class.

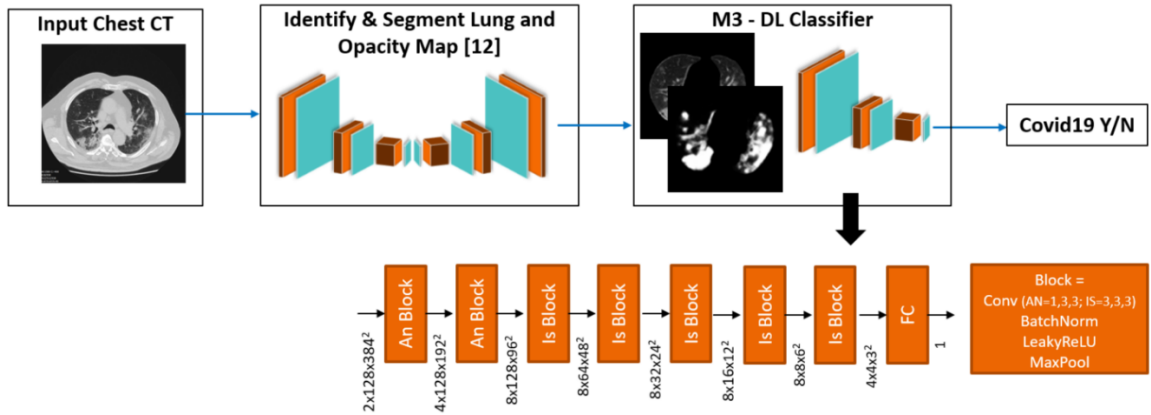

Figure 3. Heat Map of Hierarchical Clustering. This illustrates the unsupervised hierarchical clustering of the seven metrics along with cohort membership (COVID-19, other pneumonia, ILD and no pathologies) from the entire training set of 1800 cases. The metric values are standardized and rescaled to a value between 0 and 1. (a) training dataset; (b) test dataset

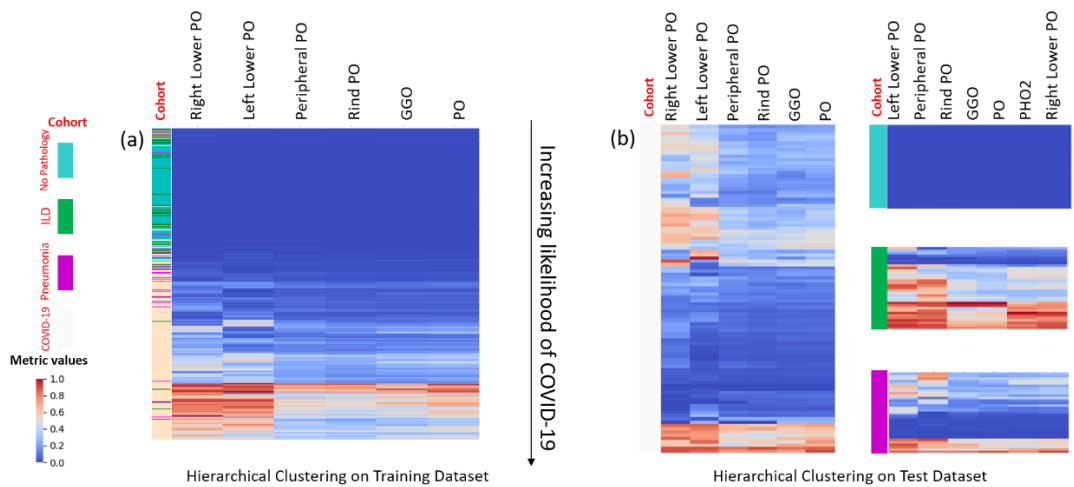

Figure 4. (a) Bootstrapped ROCs for discriminating COVID-19 from ILD, other pneumonia and no pathology control by the models proposed in this study. The models M1, M2 and M3 were evaluated with 100 COVID-19 positive, 33 ILD, 33 pneumonia and 34 healthy no pathologies CTs. The 95% confidence intervals (shown as a band) are computed by bootstrapping over 1000 samples with replacement from the predicted scores. (b) Bootstrapped ROCs for our 3D DL classifier (M3), the model proposed by Li et al (10) and the Clara model proposed by Harmon et al(15). For the model proposed by Li et al, we trained and tested on our dataset using the code provided by the authors. The 95% confidence intervals (shown as a band) are computed by bootstrapping over 1000 samples with replacement from the predicted scores.

(a)

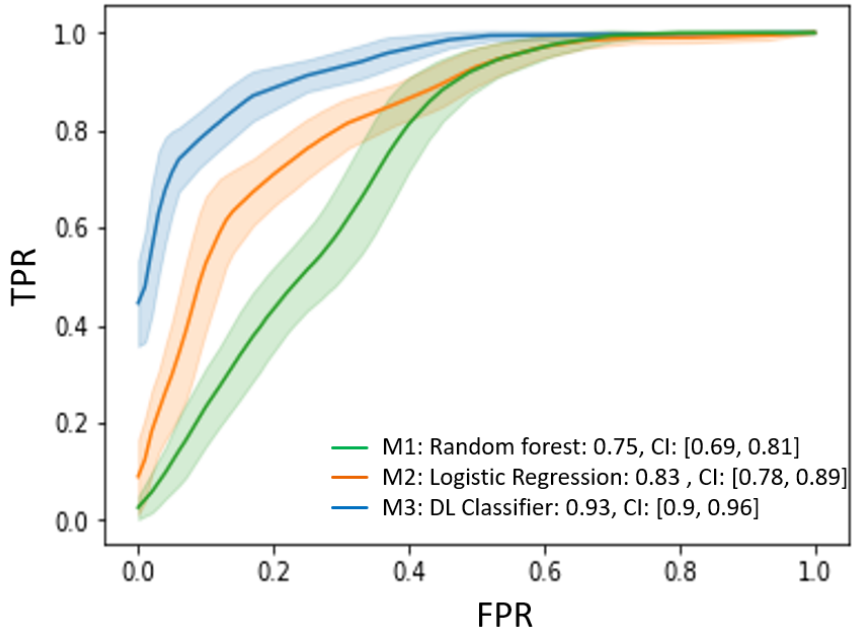

(b)

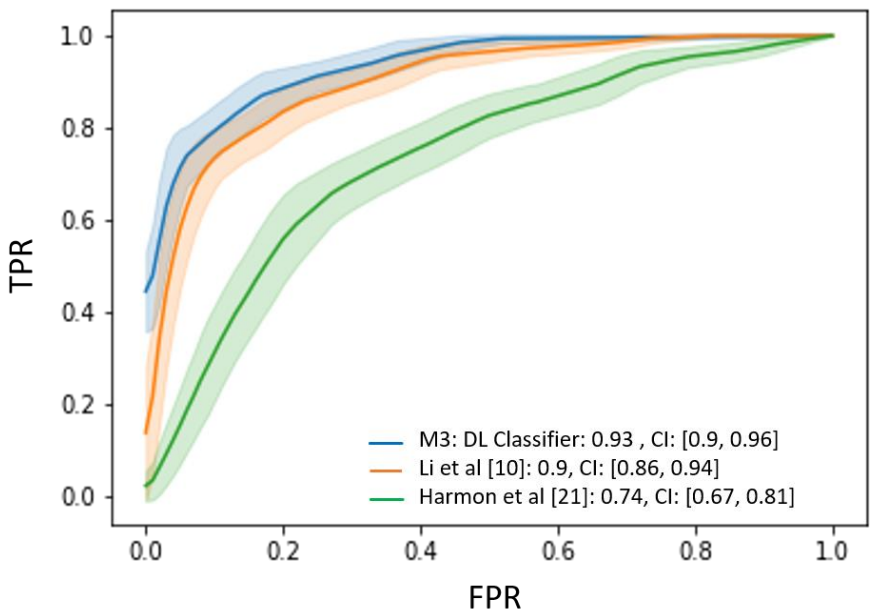

Figure 5. Examples of correctly classified COVID-19 positive patients from both methods. Red marks abnormalities associated with COVID-19. Red marks abnormalities.

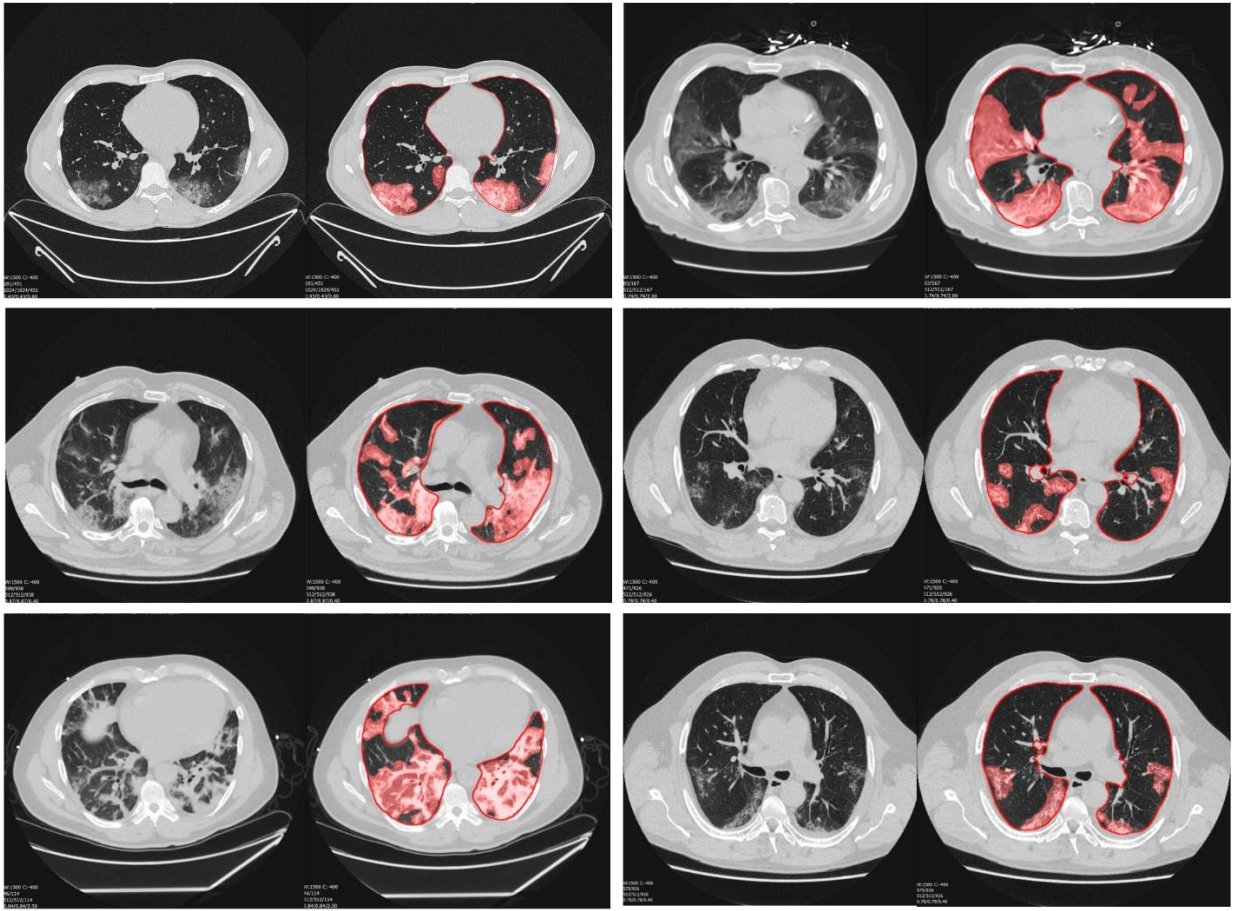

Figure 6. Examples of incorrectly classified samples by both methods: top-row COVID-19 (false negative), middle-row ILD (false positive), bottom-row other Pneumonia (false positive). Red marks abnormalities. Red marks abnormalities associated with COVID-19.

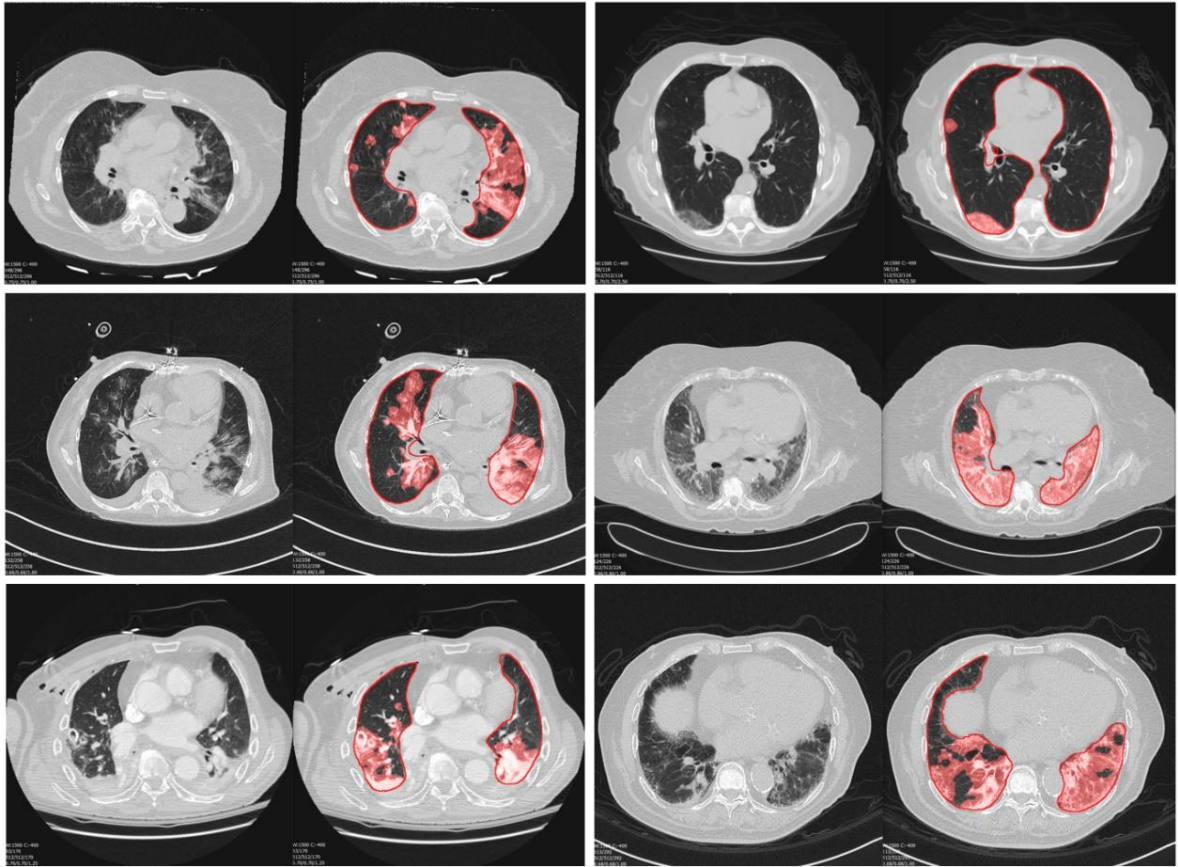

SUPLEMENTARY MATERIAL LEGENDS

Table S1 - Origins, composition and details of the Training, Validation and Test datasets

|  | COVID-19 | | |
| --- | --- | --- | --- |
| **# Data Sets** | Train: 927 | Validation: 50 | Test: 100 |
| **Data Origin** | EU: 743; North America: 184 | EU:40; North America: 10 | EU:46; North America: 54 |
| **Sex** | F:221, M:336, Unknown:370 | F:15, M:18, Unknown:17 | F:27, M:57, Unknown:16 |
| **Age** | Median:62 yrs, IQR: 51-73 | Median:61 yrs, IQR: 50-74.5 | Median:49 yrs, IQR:39-66 |

| Manufacturer | Siemens: 751; GE: 131; NMD: 18; Philips: 9; Toshiba: 15; Unknown: 3 | Siemens: 33; GE: 9; NMD: 2; Philips: 3; Toshiba: None; Unknown: 3 | Siemens: 21; GE: 35; NMD: 11; Philips: 11; Toshiba: 16; Unknown: 17 |
|---|---|---|---|
| Slice Thickness [mm] | <1.5: 802; [1.5,3]: 112; >3: 10 unknown:3 | <1.5: 39; [1.5,3]: 11; >3: None | <1.5: 41; [1.5, 3]: 48; >3: 11 |
| Reconstruction kernel | Hard: 809 Soft: 94; Unknown: 24 | Hard: 39; Soft:8; Unknown: 3 | Hard:25; Soft:57; Unknown: 18 |
| | ILD | | |
| # Data Sets | Train: 388 | Validation: 16 | Test: 33 |
| Data Origin | North America: 199 EU: 189 | North America: 16 | EU:25; North America:8 |
| Sex | F:51, M:66, Unknown:271 | F:9, M:4, Unknown:3 | F:11, M:16 Unknown:6 |
| Age | Median:58 yrs, IQR: 52-68.5 | Median:58 yrs, IQR: 50-69.25 | Median:77 yrs IQR:49-74 |
| Manufacturer | Siemens: 115; GE: 2; Unknown: 271 | Siemens: 16; | Siemens: 25 ; GE: 1; Unknown: 7 |
| Slice Thickness [mm] | <1.5: 92; [1.5,3]: 6; >3: 25; Unknown: 265 | <1.5: 14; [1.5,3]: 1; >3: 1 | <1.5: 21; [1.5, 3]: 0; >3: 7; Unknown: 5 |
| Reconstruction kernel | Hard: 73; Soft: 16; Unknown: 299 | Hard: 0; Soft:2; Unknown: 14 | Hard: 14; Soft: 4; Unknown: 15 |
| | Pneumonia | | |
| # Data Sets | Train: 189 | Validation: 16 | Test: 33 |
| Data Origin | EU: 60; North America: 129 | North America: 16 | North America: 31 EU: 2 |
| Sex | F:25, M:41, unknown:123 | F:8, M:5, Unknown:3 | F:11, M:13, Unknown:8 |
| Age | Median:55 yrs, IQR: 42.25-60 | Median:60 yrs, IQR: 50-73.75 | Median:55 yrs IQR:48.5-68.25 |
| Manufacturer | Siemens: 68; Unknown:121 | Siemens: 14 , GE:1; Unknown:1 | Siemens: 27; GE: 2; Unknown: 4 |
| Slice Thickness [mm] | <1.5: 14; [1.5,3]: 19; >3: 38; Unknown: 118 | <1.5: 13; [1.5,3]: 1; >3: 2 | <1.5: 17; [1.5, 3]: 2; >3: 1; Unknown: 13 |
| Reconstruction kernel | Hard: 28; Soft: 3; Unknown:158 | Soft: 1; Unknown:15 | Hard: 2; Soft: 0; Unknown:31 |
| | Healthy | | |
| # Data Sets | Train: 559 | Validation: 17 | Test: 34 |

| **Data Origin** | North America: 559 | North America: 17 | North America: 34 |
|---|---|---|---|
| **Sex** | F:302, M:209, Unknown:48 | F:8, M:8, Unknown:1 | F:17, M:17 |
| **Age** | Median:57 yrs, IQR: 45-66 | Median:60 yrs, IQR: 56-64 | Median:61 yrs, IQR: 56.25-65.75 |
| **Manufacturer** | Siemens: 291; GE: 184; Philips: 16; Toshiba: 7; Unknown: 61 | Siemens: 7; GE: 6; Philips: 3; Toshiba: 1 | Siemens: 11; GE: 10; Philips: 9; Toshiba: 4 |
| **Slice Thickness [mm]** | <1.5: 515; [1.5, 3]: 43; >3: 1 | <1.5: 0; [1.5, 3]: 14; >3: 3 | <1.5: 6; [1.5, 3]: 23; >3: 5 |
| **Reconstruction kernel** | Hard: 149; Soft: 19; Unknown: 391 | Hard: 4 Soft: 12, Unknown: 1 | Hard: 12 Soft: 22 |

Table S2 - Airspace Disease Severity Metrics – Definitions

Metric #1-6: Percentage of Opacity (%) or PO: The total percent volume of the lung parenchyma that is affected by the airspace disease, including ground-glass opacities and consolidations. Computed for both lungs and for each lobe.

Metric #7-12: Percentage of High Opacity (%) or PHO (>-200HU): The total percent volume of the lung parenchyma that is affected by severe disease i.e., high opacity regions including consolidation and peri-vascular thickening. High opacity is defined as the airspace disease region with mean H.U. greater than -200. Computed for both lungs and for each lobe.

Metric #13-18: Percentage of High Opacity (%) 2 (-200 to 50HU): The total percent volume of the lung parenchyma that is affected by denser airspace disease i.e., high opacity regions including consolidation. High opacity is defined as the airspace disease region with mean H.U. between -200 and 50. Computed for both lungs and for each lobe.

Metric #19: Lung severity Score (LSS): Sum of severity score for each of the five lobes. Based on PO for each lobe, severity score of a lobe is: 0 if lobe not affected, 1 if 1-25% is affected, 2 is 25-50% is affected, 3 is 50-75% is affected, 4 is 75-100% affected. (20)

Metric #20: Lung High Opacity Score (LHOS): Sum of severity score for each of the five lobes, for high opacity regions only. Based on PHO for each lobe, severity score of a lobe is: 0 if lobe not affected, 1 if 1-25% is affected, 2 is 25-50% is affected, 3 is 50-75% is affected, 4 is 75-100% affected.

Metric #21: Lung High Opacity Score (LHOS): Sum of severity score for each of the five lobes, for high opacity regions excluding vasculature (threshold 50 HU). Based on PHO for each lobe, severity score of a lobe is: 0 if lobe not affected, 1 if 1-25% is affected, 2 is 25-50% is affected, 3 is 50-75% is affected, 4 is 75-100% affected.

Metric #22: Bilaterality: True if both right and left lungs are involved, false if only one of the two or none is involved.

Metric #23: Number of Affected Lobes: Number of lobes affected by the disease.

Metric #24: Number of Total Lesions: Number of affected regions in the lung.

Metric #25: Number of Peripheral Lesions: Number of lesions that are in the periphery of the lung. Not including apex and mediastinal regions. See Fig S1(a). Any abnormality that intersects with the peripheral border is considered a peripheral lesion. (16)

Metric #26: Number of Lesions in the Rind: Number of regions that are in the rind of the lung as defined in (17) (See Fig S1(b)). Any abnormality that intersects with the "rind" is considered a lesion in the rind.

Metric #27: Number of Lesions in the Core: Number of regions that are in the core of the lung as defined in (17) (See Fig S1(b)). Any abnormality that does not intersect with the rind, is considered a core lesion.

Metric #28: Percent of Peripheral Distribution: Given by the number of peripheral lesions divided by the number of total lesions.

Metric #29: Percent of Peripheral Lesions: The total percent volume of the lung parenchyma that is affected by disease for peripheral lesions only.

Metric #30: Percent of Rind Lesions: The total percent volume of the lung parenchyma that is affected by disease for rind lesions only.

Metric #31: Percent of Core Lesions: The total percent volume of the lung parenchyma that is affected by disease for core lesions only.

Metric #32: Percentage of Ground Glass Opacity: The total percent volume of the lung parenchyma that is affected by less dense airspace disease i.e., lesions which are characterized as GGO only. GGO is defined as the airspace disease region with mean H.U. less than -200.

Figure S1 - Example of core and rind partition for CT lung segmentation. (a) Shows the periphery region. This definition excludes the apex and mediastinal border. (b) Shows the core and rind regions.

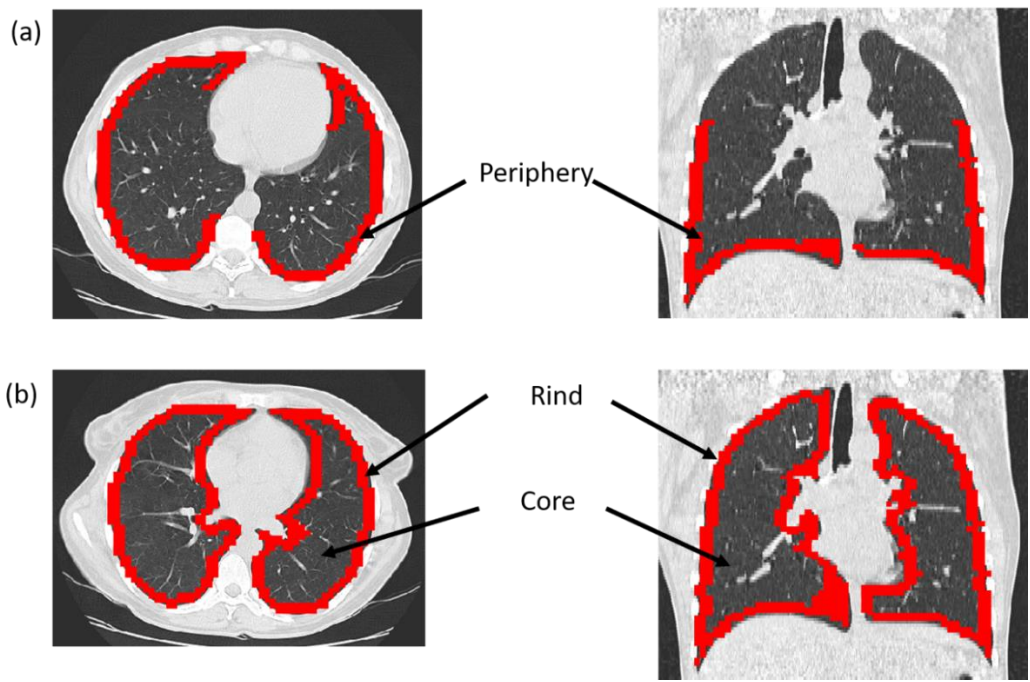

Figure S2 - Examples of correctly classified negative samples from both methods – top row ILD, bottom row other Pneumonia.

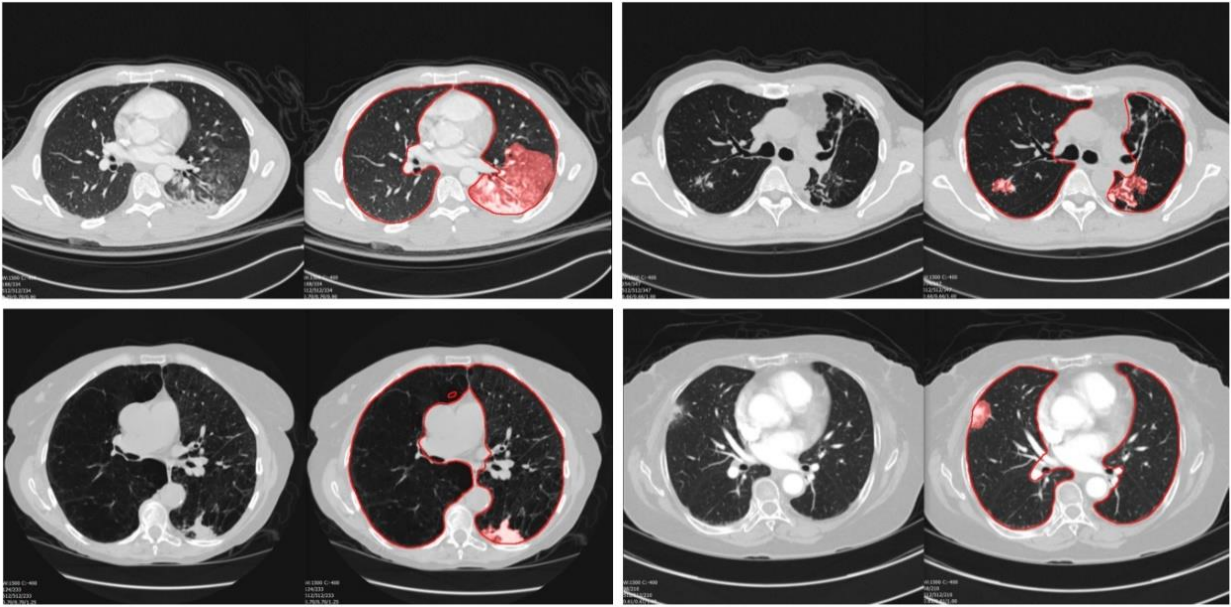

Figure S3 - Examples of samples correctly classified by the DL classifier but incorrectly classified by the metric-based classifier.

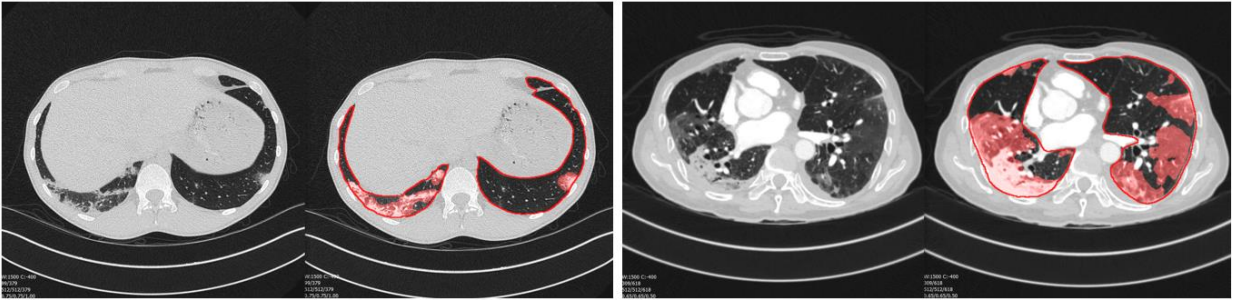